# THE NARROW-BAND RESONANCE IN TWO-LAYER METALLIC ROUND TUBE

M. Ivanyan, V. Tsakanov, A. Grigoryan, A. Tsakanian

The regularities of narrow-band resonance in a two-layer circular waveguide due to the radiation of the bunch, moving along its axis are investigated. Its connection with the synchronous component of a two-layer $TM_{01}$ mode of the waveguide is established. Also performed numerical and analytical calculations of wake functions (particle radiation fields) and are investigated its properties. The possibility of using a two-layer waveguide as a powerful source of quasi-monochromatic terahertz radiation and as an efficient device for the two-beam acceleration is shown. Preliminary experimental scheme is proposed.

## 1. Introduction

Development of high-power sources of coherent radiation in the terahertz frequency range is now an urgent task. The most high-profile projects, such as a Eurorean XFEL [1] or SvissFEL [2] are based on the principle of a free electron laser, suitable for use up to X-ray frequencies. Recently, intensive research on the development of these sources on the basis of the Cherenkov radiation produced by the passage of the sequence of bunches of charged particles in dielectric [3] and corrugated [4] waveguides are performing. Theoretical studies in this direction are also held in CANDLE [5]. Here we are examining the use of two-layer metal waveguide as a narrow-band radiation source and as a devise for two-beam acceleration. Studies are based on the phenomenon, which was previously identified: the emergence of a narrow-band resonance in a circular waveguide with a high conductivity of wall's metal and with the thin low-conducting metalic layer inside [6]. For the practical application of this phenomenon, its comprehensive study in order to clarify its nature, is necessary. Numerical calculations indicate on the rigid dependence of the resonance frequency of the radius of the waveguide and of the thickness of the inner layer. This allows to adjustment of the resonant frequency with the help of selection of geometrical parameters of the waveguide.

We present numerical examples, calculated by the exact formulae [7], showing the laws of appearance of the resonance and its movements along the frequency scale. Nature of resonance is obtained by establishing a correspondence between the resonance and extreme properties of the frequency distribution of the fundamental TM mode of the waveguide. The analytical expressions for the resonant frequency, linking it with geometric and electro-dynamic parameters of the waveguide, are obtained.

Thus, the resonance radiation is due to the mode $TM_{01}$ of the waveguide, which allows to generate it, simultaneously suppressing the others, with the help of corresponding sequence of bunches. This sequence of bunches, emitting simultaneously, enhance the resonance effect.

A notable feature of dielectric waveguides is possibility to use them as a means of two-beam acceleration. We show that the considered two-layer metallic waveguide in the presence therein of synchronous (resonant) mode can be used as an effective device for the two-beam acceleration as well.

The presence in this waveguide, in contrast to a dielectric waveguide, the single resonance of longitudinal impedance, in which concentrated most of the energy in the vicinity of the resonance frequency, makes it for using both as an effective source of coherent radiation, and as an effective device for two-beam acceleration.

The paper is organized as follows: in the second chapter describe the exact analytical and asymptotic expressions for the longitudinal impedance of a two-layer metal pipe. The third chapter is devoted to the numerical analysis of the longitudinal impedance of a two-layer tube to identify of its key properties. In the fourth chapter, asymptotic expressions for the impedance of a two-layer tube, allow to obtain analytical expressions for the resonant frequencies. are derived. In the fifth chapter of the correspondence between the resonant frequency and the frequency at which there is a synchronization of the phase velocity of the waveguide eigenmode $TM_{01}$ with the particle velocity. In the sixth chapter the wakefield radiation is calculated and its properties is defined. In the seventh and eighth chapters the possibility of using a two-layer metal waveguide as a source of terahertz radiation and as a device for two-beam acceleration is shown.

## 2. Longitudinal impedance of two-layer tube

The formulae for the two-layer round tube (Fig.1) longitudinal impedance calculated with varying degrees of accuracy are given in several papers. For numerical calculations we are using the ultrarelativistic limit of the exact expression (valid for arbitrary frequencies) obtained on the basis of the matrix method described in [7]:

$$Z_\| = j\frac{Z_0}{\pi k a_1^2}\left(1 - \frac{2}{ka_1}Q\right)^{-1}. \tag{1}$$

Here $Z_0 = 120\pi$ Ω impedance of free space, $k = \omega/c$ wavenumber ($\omega$ frequency, $c$ speed of light), $a_1$ inner radius of the tube and

$$Q = \frac{\varepsilon_1 k}{\varepsilon_0 \chi_1} \frac{\varepsilon_1 \chi_2 U_4^{(1)} U_1^{(2)} - \varepsilon_2 \chi_1 U_3^{(1)} U_3^{(2)}}{\varepsilon_1 \chi_2 U_2^{(1)} U_1^{(2)} - \varepsilon_2 \chi_1 U_1^{(1)} U_3^{(2)}} \tag{2}$$

with

$$\begin{aligned}
U_1^{(i)} &= K_0(\chi_i a_{i+1})I_0(\chi_i a_i) - I_0(\chi_i a_{i+1})K_0(\chi_i a_i) \\
U_2^{(i)} &= K_0(\chi_i a_{i+1})I_0'(\chi_i a_i) - I_0(\chi_i a_{i+1})K_0'(\chi_i a_i) \\
U_3^{(i)} &= K_0'(\chi_i a_{i+1})I_0(\chi_i a_i) - I_0'(\chi_i a_{i+1})K_0(\chi_i a_i) \\
U_4^{(i)} &= K_0'(\chi_i a_{i+1})I_0'(\chi_i a_i) - I_0'(\chi_i a_{i+1})K_0'(\chi_i a_i)
\end{aligned} \tag{3}$$

and $I_0(x), K_0(x)$ the modified Bessel functions of first and second kind and zero order. Electrodynamic properties of tube (for non-magnetic materials) described by the electrical permeabilities and transverse propagation constants of inner $(\varepsilon_1, \chi_1)$ and outer $(\varepsilon_2, \chi_2)$ metallic layers:

$$\varepsilon_{1,2} = \varepsilon_0 + j\frac{\sigma_{1,2}}{\omega}, \quad \chi_{1,2} = \sqrt{-j\sigma_{1,2}\mu_0\omega}, \quad \mathrm{Im}\,\chi_{1,2} \geq 0 \tag{4}$$

with $\sigma_{1,2}$ conductivities of layer's metal, $\varepsilon_0$, $\mu_0$ electric and magnetic constants of vacuum.

The expressions (1), (2) turn to the formulae for the single layer tube under condition $a_3 = a_2$. Under this condition $U_1^{(2)} = U_4^{(2)} = 0$, $U_2^{(2)} = (\chi_2 a_2)^{-1}, U_3^{(2)} = -(\chi_2 a_2)^{-1}$ and one obtains the impedance for single-layer tube with inner and outer radii $a_1$ and $a_2$ respectively:

$$Z_\| = j \frac{Z_0}{\pi k a_1^2} \left(1 + \frac{2}{\chi_1 a_1} \frac{\varepsilon_1}{\varepsilon_0} \frac{U_3^{(1)}}{U_1^{(1)}}\right)^{-1}. \tag{5}$$

In opposite case of $a_1 = a_2$ we have the impedance of tube with inner and outer radii $a_2$ and $a_3$:

$$Z_\| = j \frac{Z_0}{\pi k a_2^2} \left(1 + \frac{2}{\chi_2 a_2} \frac{\varepsilon_2}{\varepsilon_0} \frac{U_3^{(2)}}{U_1^{(2)}}\right)^{-1} \tag{6}$$

Finally, if we substitute in (2) $\varepsilon_1 = \varepsilon_2$, we obtain the impedance of the waveguide with the inner and outer radii $a_1$ and $a_3$, respectively:

$$Z_\| = j \frac{Z_0}{\pi k a_1^2} \left(1 + \frac{2}{\chi_1 a_1} \frac{\varepsilon_1}{\varepsilon_0} \frac{\tilde{U}_3}{\tilde{U}_1}\right)^{-1}, \tag{7}$$

where

$$\tilde{U}_1 = I_0(\chi_1 a_1) K_0(\chi_1 a_3) - I_0(\chi_1 a_1) K_0(\chi_1 a_3)$$
$$\tilde{U}_3 = I_0'(\chi_1 a_1) K_0(\chi_1 a_3) - I_0(\chi_1 a_1) K_0'(\chi_1 a_3) \tag{8}$$

Thus, the expression for the impedance (1), (2) is self-consistent and does not lead to contradictions.
For the theoretical investigations and modeling we are using the simplified expression, developed in works [6,7] for two-layer tube longitudinal impedance, which are valid for not too low frequencies $(\min(\chi_{1,2} a_{1,2}) \gg 1)$:

$$Z_\|^0 = j \frac{Z_0}{\pi k a_1^2} \left(1 + \frac{2}{a_1} \frac{\varepsilon_1}{\varepsilon_0 \chi_1} \frac{\varepsilon_1 \chi_2 th(\chi_2 d_2) th(\chi_1 d_1) + \varepsilon_2 \chi_1}{\varepsilon_1 \chi_2 th(\chi_2 d_2) + th(\chi_1 d_1) \varepsilon_2 \chi_1}\right)^{-1}, \tag{9}$$

where $d_{1,2}$ thicknesses of inner and outer layers respectively. This expression is self-consistent as well.

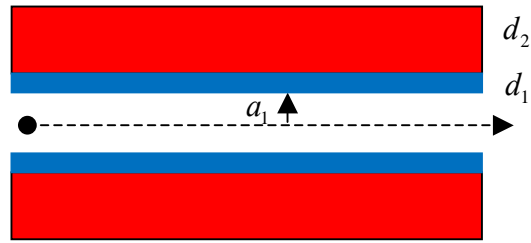

Fig.1. Two-layer tube

## 3. Numerical investigations

The dependence of the resonance position on the frequency scale of the thickness of the inner layer (for a fixed conductivity) is shown in [7]. Now we need to examine the laws of appearance the resonance more detail. We are also interested in changing the laws of resonance and its displacements along the frequency scale depending not only on the thickness of the inner layer, but on the conductivities of both of layers and from the inner radius of waveguide as well. Figure 2 shows the resonance distribution for different thicknesses of the lower layer and for two different values of conductivities of external layer. Figure 3 shows the distribution of the resonance curves for different values of the inner radii of the waveguide with identical values of external and internal layer conductivities. Graphs in the left side of Figure 2 and 3 allow us to compare the resonance properties of the waveguide with different conductivities of inner layer at a constant value of the inner radius of the waveguide. Finally, Figure 4 shows the distribution of the impedance in the case of high conductivity of lower layer in comparison with the upper layer conductivity.

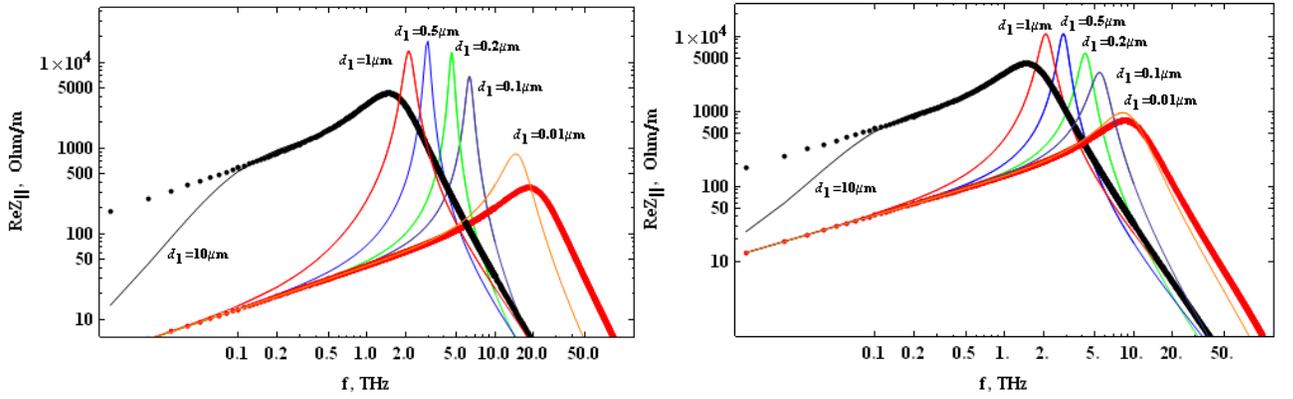

Fig.2. Real part of longitudinal impedance of two-layer tube; $a_1 = a_2 - d_1$, $a_2 = 1mm$, $a_3 = 2mm$; $\sigma_1 = 3 \cdot 10^4 \ \Omega^{-1} m^{-1}$, $\sigma_2 = 58 \cdot 10^6 \ \Omega^{-1} m^{-1}$ (left), $\sigma_2 = 58 \cdot 10^5 \ \Omega^{-1} m^{-1}$ (right).

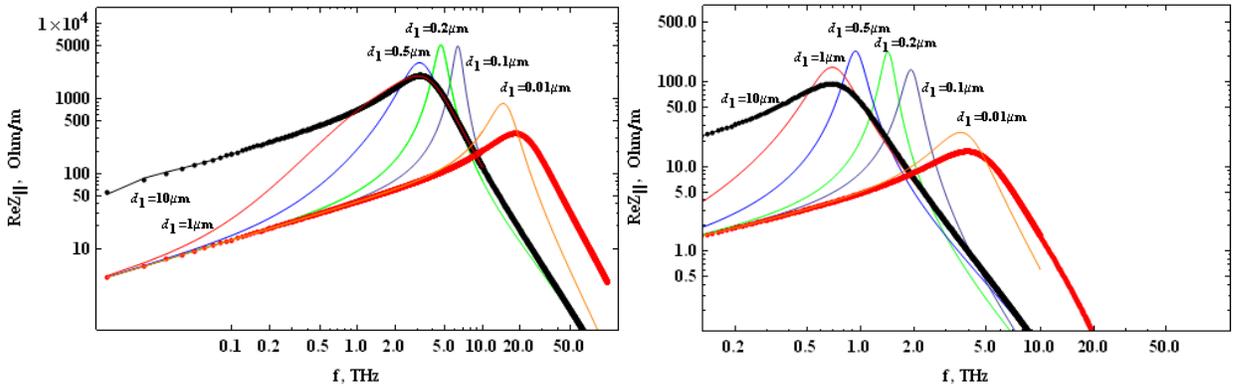

Fig.3. Real part of longitudinal impedance of two-layer tube; $\sigma_1 = 3 \cdot 10^5 \ \Omega^{-1} m^{-1}$, $\sigma_2 = 58 \cdot 10^6 \ \Omega^{-1} m^{-1}$, $a_1 = a_2 - d_1$; $a_2 = 1mm$, $a_3 = 2mm$ (left); $a_2 = 1cm$, $a_3 = 2cm$ (right).

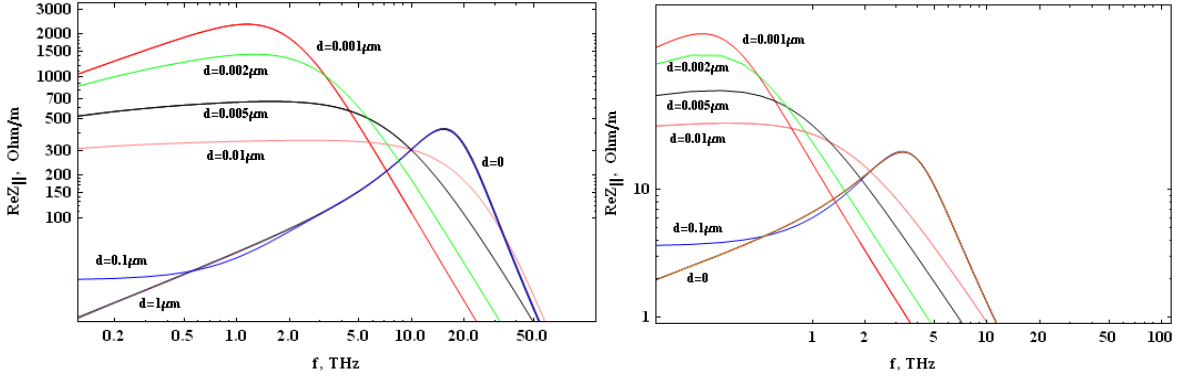

Fig.4. Real part of longitudinal impedance of two-layer tube; $\sigma_2 = 3 \cdot 10^4\ \Omega^{-1}m^{-1}$, $\sigma_1 = 58 \cdot 10^6\ \Omega^{-1}m^{-1}$, $a_1 = a_2 - d$, $d = d_1$; $a_2 = 1mm$, $a_3 = 2mm$ (left); $a_2 = 1cm$, $a_3 = 2cm$ (right).

Figure 4 shows at once that for the high conductivity of the lower layer (in comparison with the outer layer, $\sigma_1 > \sigma_2$) conditions for resonance are not met. Impedance distribution is a smooth curve. At low frequencies in the case of thin inner coating, it is close to the impedance of the tube, made entirely of the material of the top layer. Conversely, at high frequencies and in the case of the relatively thick inner layer, the impedance curve approaches the shape of the curve of impedance of a tube, made entirely from the metal of inner cover. These limiting cases can be explained by the corresponding skin-depth. At low frequencies, the thickness of the skin layer can greatly exceed the thickness of the inner coating and the field of particle at these frequencies interacts mainly with the outer layer of the wall of the waveguide. The higher the frequency, the smaller the depth of the skin layer and at high frequencies, the particle field interacts mainly with the lower layer of the waveguide wall, without penetrating the top layer. At intermediate frequencies we have the mutual suppression of the combined interaction in antiphase of both layers.

As follows from Figures 2 and 3, the resonance (synchronous interaction) between two layers occurs in the case of $\sigma_1 < \sigma_2$. The frequency range in which the resonance can occur, is between $k_1 \approx 1.74/s_1$ and $k_2 \approx 1.74/s_2$ ($s_{1,2} = \left(2a_{1,2}^2 \varepsilon_0 c / \sigma_{1,2}\right)^{1/3}$ is the characteristic distance of corresponding tube [8]), which correspond to the maximum values of the real part of impedance of the tube made entirely from the material of external and internal layer. Reducing the thickness of the inner cover increases the resonance frequency. But too thin coating, as well as too thick, prevents arising of resonance. In the first case, the particle it just does not feel of inner cover, in the second case it only interacts with the inner coating. Thus, if the thickness of the lower layer is much greater than the skin depth $\delta_1$ at the frequency of $\omega_1$ or much smaller than the skin depth $\delta_2$ at the frequency of $\omega_2$, the resonance will not occur. The necessary condition for the resonance is $\delta_2 < d_1 < \delta_1$.

We now consider the dependence of the resonance properties of the waveguide from the main electrodynamic and geometrical parameters. Comparison of the left and right part of Figure 2 indicates on the increase the level of resonance, on its shift towards the higher frequencies and on the narrowing of the resonance curve with increasing conductivity of the upper layer of the wall of the waveguide.

The main parameters of the resonance curves are shown in Table 1. In columns I (III) and II (IV) of Table the frequency at which resonance occurs $\left(f_{\max}, THz\right)$, the level $\left(Z_{\|,\max}, \Omega/m\right)$, geometrical resonant frequency $(f_d, THz)$, resonant frequency with the material correction $(f'_d, THz)$ and width $(\Delta f, THz)$ of the resonance curve (calculated for the half-level) depending on the thickness of the inner coating $(d_1, \mu m)$ for the left and right parts of Figure 2 are shown.

Table 1. The main parameters of the resonance curves shown in Figure 2 (left, I; right, II) and the left (III) and right (IV) sides of Figure 3 (III).

| $d_1$ | $f_{max}$ | $Z_{\parallel,max}$ | $f_d$ | $f'_d$ | $\Delta f$ | $f_{max}$ | $Z_{\parallel,max}$ | $f_d$ | $f'_d$ | $\Delta f$ |
|---|---|---|---|---|---|---|---|---|---|---|
| I. $\sigma_1 = 3\cdot 10^4\ \Omega^{-1}m^{-1}$, $\sigma_2 = 58\cdot 10^6 \Omega^{-1}m^{-1}$ | | | | | | III. $\sigma_1 = 3\cdot 10^5\ \Omega^{-1}m^{-1}$, $\sigma_2 = 58\cdot 10^6 \Omega^{-1}m^{-1}$ | | | | |
| $a_1 = 1mm$ | | | | | | $a_1 = 1mm$ | | | | |
| 10 | 1.48 | 4337 | 0.6758 | 0.6738 | 1.56 | 3.19 | 2015 | 0.6752 | 0.6738 | 3.38 |
| 1 | 2.12 | 13589 | 2.1352 | 2,11 | 0.4 | 3.05 | 1997 | 2.13 | 2.11 | 3.42 |
| 0.5 | 2.97 | 18377 | 3.02 | 2.96 | 0,3 | 3.14 | 3012 | 3.019 | 2.96 | 1.78 |
| 0.2 | 4.6 | 12953 | 4.775 | 4.59 | 0.46 | 4.65 | 5220 | 4.77 | 4.59 | 1.0 |
| 0.1 | 6.32 | 6834 | 6.7523 | 6.32 | 0.86 | 6.36 | 5015 | 6.75 | 6.32 | 1.18 |
| 0.01 | 14,5 | 851 | 21,35 | 13.7 | 7.8 | 14.6 | 862 | 21 | 14 | 7.6 |
| II, $\sigma_1 = 3\cdot 10^4\ \Omega^{-1}m^{-1}$, $\sigma_2 = 58\cdot 10^5\ \Omega^{-1}m^{-1}$ | | | | | | IV. $\sigma_1 = 3\cdot 10^5\ \Omega^{-1}m^{-1}$, $\sigma_2 = 58\cdot 10^6 \Omega^{-1}m^{-1}$ | | | | |
| $a_1 = 1mm$ | | | | | | $a_1 = 1cm$ | | | | |
| 10 | 1.48 | 4336 | 0.6752 | 0.671 | 1.56 | 0.63 | 93 | 0.2135 | 0.2137 | 0.72 |
| 1 | 2.08 | 10855 | 2.13 | 2.06 | 0.5 | 0.69 | 147 | 0.6752 | 0.6616 | 0.34 |
| 0.5 | 2.85 | 10904 | 3.02 | 2.84 | 0.52 | 0.94 | 228 | 0.95 | 0.92 | 0.24 |
| 0.2 | 4.21 | 5965 | 4.77 | 4.20 | 1.02 | 1.42 | 224 | 1.51 | 1.41 | 0.26 |
| 0.1 | 5.44 | 3300 | 6.75 | 5.39 | 1.94 | 1.9 | 138 | 2.13 | 1.89 | 0.42 |
| 0.01 | 8.22 | 961 | 21.35 | -2.8 | 7.3 | 3.93 | 15 | 0.0067 | 0.067 | 4.34 |

As follows from a comparison of the left side of Figure 2 and 3, as well as from the comparison of columns I and III of Table 1, increasing the conductivity of the material of the lower layer 10 times reduces the levels of resonance, leads to their broadening, but does not affect significantly on the value of the resonant frequencies of the main peaks (above the level of normal maximum for copper pipe with $\sigma = 58\cdot 10^6 \Omega^{-1}m^{-1}$) for the $d_1 = 0.2\mu m$ and $d_1 = 0.1\mu m$.

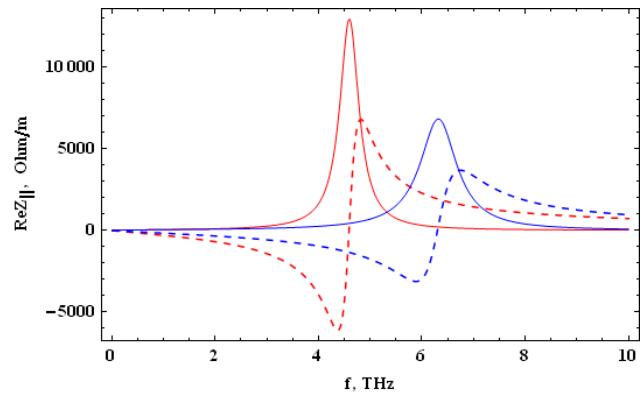

Fig.5. Real (solid) and imaginary (dotted) part of longitudinal impedance of two-layer tube; $\sigma_1 = 3\cdot 10^4\ \Omega^{-1}m^{-1}$, $\sigma_2 = 58\cdot 10^6\ \Omega^{-1}m^{-1}$, $a_1 = a_2 - d_1$; $a_2 = 1mm$, $a_3 = 2mm$; $d_1 = 0.1\mu m$ (blue), $d_1 = 0.2\mu m$ (red).

Finally, the effects of increasing the inner radius of the waveguide can be detected by comparing the left and right side of Figure 3. The increase of the inner radius leads to lower level of resonance and to reduction of resonance frequency.

In conclusion we note that in each case, shown in Figures 2 and 3 (ie, each combination of values of the conductivity of the upper and the lower layer and the inner radius of the waveguide), corresponds to the certain thickness of the inner layer, for which the maximum amount of resonance is obtained.

For a more complete understanding of the nature of obtained resonance it is necessary also investigate the behavior of the imaginary component of the impedance of a two-layer pipe. It is especially important to look of its behavior in the vicinity of the resonance frequency (Fig.5).

It's seen (Fig. 5) that the imaginary part of the impedance is close to zero in the vicinity of the resonance frequency. Moreover, the resonant frequency and the zero value of the imaginary part converge with increasing conductivity of the material of the upper layer and tend to a definite finite limit (Table 2). Table 2 shows the values of the resonance frequencies $(f_{rez}, THz)$ and the frequencies corresponding to zero values of the imaginary part of the impedance $(f_{zero}, THz)$ for three different values of the conductivity $(\sigma_2, \Omega^{-1}m^{-1})$ of the upper layer and their difference $(\Delta f = f_{rez} - f_{zero}, THz)$. For comparison are also given the same parameters for a single-layered tube (Fig. 6). As can be seen from the Table, as opposed to a two-layer tube, single-layer tube frequency difference $\Delta f$ increases with increasing the conductivity of the material of the external layer.

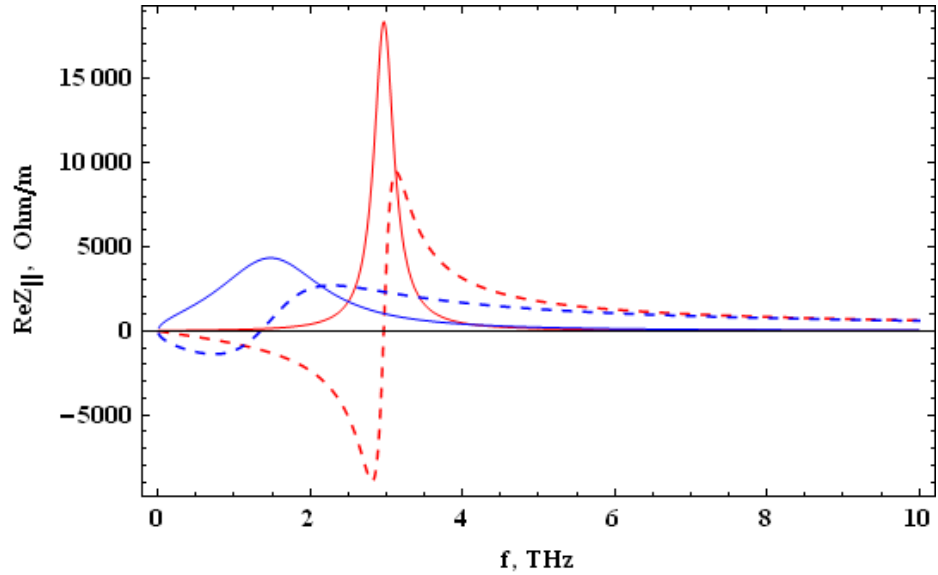

Fig. 6. Real (solid) and imaginary (dotted) part of longitudinal impedance of two-layer tube, $\sigma_1 = 3 \cdot 10^4 \, \Omega^{-1}m^{-1}$, $\sigma_2 = 58 \cdot 10^6 \, \Omega^{-1}m^{-1}$, $a_2 = 1mm$, $d_1 = 0.5\mu m$ (red) and for the single-layer tube with $\sigma = 3 \cdot 10^4 \, \Omega^{-1}m^{-1}$ and inner radius $1mm$ (black).

The data in Table 2 indicate the different mechanisms of narrow-band resonance in two-layer tube and broad-band resonance in single-layer pipe.

Proximity to zero of the imaginary part of the impedance at the resonant frequency for the real part indicates a synchronous mode in a set of fundamental solutions (eigenmodes) of double-layer waveguide. This situation occurs under certain conditions, namely, the small thickness and low conductivity of the lower layer. However, the thickness of the lower layer has a certain lower limit at which there is still a narrow-band resonance. With further thinning of the lower layer the resonance curve broadens and transforms into a broadband resonance for the single-layer pipe.

Table 2. Comparison of resonance parameters for a two-layer and single-layer pipe; $a_1 = 1mm$, $d_1 = 0.2\mu m$, $\sigma_1 = 3\cdot 10^4 \Omega^{-1} m^{-1}$.

| $\sigma_2, \Omega^{-1}m^{-1}$ | Two-layer tube, *THz* | | | Single-layer tube, *THz* | | |
|---|---|---|---|---|---|---|
| | $f_{rez}$ | $f_{zero}$ | $\Delta f$ | $f_{max}$ | $f_{zero}$ | $\Delta f$ |
| $58 \cdot 10^5$ | 4.21 | 4.18 | 0.03 | 7.07 | 6.44 | 0.63 |
| $58 \cdot 10^6$ | 4.6 | 4.59 | 0.01 | 15.24 | 13.89 | 1.35 |
| $58 \cdot 10^7$ | 4.72 | 4.72 | 0.00 | 32.81 | 29.8 | 3.01 |

In the following paragraphs we will try to obtain (in the simplest cases) the analytical expressions for the resonant frequency and correlate them with the natural frequencies of the main TM mode of two-layer waveguide.

## 4. Analytical treatment

Let's simplify the expression (9) for the impedance, setting $d_2$ large enough $(d_2 \to \infty)$. In this case $th(\chi_2 d_2) \to \infty$ and one can write instead of (9):

$$Z_\parallel^0 = j\frac{Z_0}{\pi k a_1^2}\left(1 + \frac{2}{a_1}\frac{\varepsilon_1}{\varepsilon_0 \chi_1}\frac{\varepsilon_1 \chi_2 \, th(\chi_1 d_1) + \varepsilon_2 \chi_1}{\varepsilon_1 \chi_2 + th(\chi_1 d_1)\varepsilon_2 \chi_1}\right)^{-1} \qquad (10)$$

Imposing the usual restrictions on the very high frequencies:

$$\omega \ll \sigma_1/\varepsilon_0 \qquad (11)$$

and assuming $\sigma_2 \gg \sigma_1$, expression (10) can be simplified to the form

$$Z_\parallel^0 = j\frac{Z_0}{\pi k a_1^2}\left(1 + \frac{2}{a_1}\frac{\varepsilon_1}{\varepsilon_0 \chi_1}\frac{S\, th(\chi_1 d_1) + 1}{S + th(\chi_1 d_1)}\right)^{-1} \qquad (12)$$

For small $S = \varepsilon_1 \chi_2 / \varepsilon_2 \chi_1 \approx \sqrt{\sigma_1/\sigma_2}$ it can be written in the form

$$Z_\parallel^0 = j\frac{Z_0}{\pi k a_1^2}\left(1 + \frac{2}{a_1}\frac{\varepsilon_1}{\varepsilon_0 \chi_1}\left(\frac{1}{th(\chi_1 d_1)} - \frac{S}{sh^2(\chi_1 d_1)}\right)\right)^{-1} \qquad (13)$$

Finally, for small $d_1$, we have

$$Z_\parallel^0(\omega) = j\frac{Z_0 c}{\pi a_1^2}(X + jY)^{-1} \qquad (14)$$

$$X(\omega) = \omega - \frac{2c^2}{a_1 d_1 \omega} + S \frac{\sqrt{2\mu_0 \sigma_1 \omega}}{a_1 d_1^2 \omega \sigma_1 \mu_0^2} \left( \frac{1}{\varepsilon_0 \omega} + \frac{1}{\sigma_1} \right), \quad Y(\omega) = \frac{2c^2 \varepsilon_0}{a_1 d_1 \sigma_1} + S \frac{\sqrt{2\mu_0 \sigma_1 \omega}}{a_1 d_1^2 \omega \sigma_1 \mu_0^2} \left( \frac{1}{\varepsilon_0 \omega} - \frac{1}{\sigma_1} \right). \tag{15}$$

At taking into account condition (11) we can neglect the second term in brackets (15):

$$X(\omega) = \omega - \frac{\omega_d^2}{\omega} + S \frac{\alpha}{\omega^{3/2}}, \quad Y(\omega) = \beta + S \frac{\alpha}{\omega^{3/2}} \tag{16}$$

Here

$$\omega_d = c\sqrt{2/a_1 d_1}, \quad \alpha = \frac{a_1 \omega_d^4 \sqrt{2\varepsilon_0}}{4c\sqrt{\sigma_1}}, \quad \beta = \omega_d^2 \frac{\varepsilon_0}{\sigma_1}. \tag{17}$$

Imaginary component of the impedance vanishes at $X = 0$. If the outer layer is a perfectly conducting ($S = 0$), the solution of this equation is $\omega = \omega_d$ and is expressed by the main geometrical parameters of the waveguide: its inner radius and thickness of the inner layer. At high but finite conductivity of the outer layer ($S \ll 1$) the solution can be approximately written as:

$$\omega_d' = \omega_d - \omega_d \tilde{\kappa}_d^{3/2}/8, \tag{18}$$

where $\tilde{\kappa}_d = \frac{\omega_d}{c} s_2$ dimensionless wavenumber and

$$s_2 = \left( 2c\varepsilon_0 a_1^2 / \sigma_2 \right)^{1/3} \tag{19}$$

is the characteristic length of outer layer.
The real part of impedance may be written as:

$$\operatorname{Re} Z_\parallel^0 = \frac{Z_0 c}{\pi a_1^2} \frac{Y(\omega)}{X^2(\omega) + Y^2(\omega)} \tag{20}$$

In particular, for the perfectly conducting external layer one obtains:

$$\operatorname{Re} Z_\parallel^0 = \frac{Z_0 c}{\pi a_1^2} \frac{\beta \omega^2}{\left( \omega^2 - \omega_d^2 \right)^2 + \beta^2 \omega^2} \tag{21}$$

It is easy to show that this expression has a maximum at the same frequency $\omega = \omega_d = c\sqrt{2/a_1 d_1}$, where vanishes the imaginary part of the impedance at a perfectly conducting outer layer. We can also show that in the case of non-zero, but small $S$ ($S \ll 1$) under condition of $\beta \ll \omega_d$, the maximum of the real part of the impedance, ie, the resonant frequency is achieved simultaneously with the vanishing of its imaginary part (see (18)). Thus, for the perfectly conducting outer layer the resonant frequency $\omega_d$ is determined by the geometrical dimensions of waveguide. The allowance for the finite conductivity of the outer layer is reduced in a first approximation to the material supplements containing the conductivity of the layer. Table 1 shows the resonance frequencies, calculated by the numerical method, the geometric frequency and frequencies obtained with the material additives. Identical or similar results are marked in red. As the table shows, there are a satisfactory agreement for the main resonance curves.

Thus, the numerical calculations and theoretical studies as well indicate the presence of resonant modes in a two-layer waveguide with high conductivity of the upper layer of the wall and with the low conductivity of inner layer. We now proceed to establish the correspondence between resonance in the two-layer waveguide and the extreme properties of its fundamental TM mode.

### 5. Two layer waveguide transverse eigenvalue calculation. Surface impedance.

The equation for the eigenvalues of the fundamental solutions of homogeneous Maxwell's equations in two-layer waveguide can be obtained through the use of cross-linking procedure for solutions in each of the layers of the wall of the waveguide and the inner and outer regions on the borders, that separates layers in the wall of the waveguide. Without going into details, we write the final result for the axisymmetric TM modes for unbonded waveguide $(a_3 \to \infty)$:

$$\frac{1}{v_{0,i} a_1} \frac{J_1(v_{0,i} a_1)}{J_0(v_{0,i} a_1)} = F \tag{22}$$

where $J_{0,1}(x)$ are the Bessel functions of first kind and zero and first order, $v_{0,i}$ the transverse eigenvalues $(i = 1,2,3...)$ in the inner vacuum region $(0 \le r \le a_1)$, and

$$F = -\frac{\varepsilon_1}{\varepsilon_0 v_{1,i} a_1} \frac{tg(v_{1,i} d_1) S - j}{S + j tg(v_{1,i} d_1)}. \tag{23}$$

Here, as before, $S \approx \sqrt{\sigma_1/\sigma_2}$ and $v_{1,i}^2 = k^2(\varepsilon_1 \mu_1/\varepsilon_0 \mu_0 - 1) + v_{0,i}^2$ the transverse eigenvalues $(i = 1,2,3...)$ in the inner layer region $(a_1 \le r \le a_2 = a_1 + d_1)$. For non-magnetic material of inner layer $(\mu_1 = \mu_0)$ and for not very small frequencies and not very large number of eigenvalue $v_{0,i}$ one can write:

$$v_{1,i} = \sqrt{-jk\sigma_1 Z_0} = v_1, \tag{24}$$

in the form, independent from the eigenvalue number. This allows us to deduce from (22) the expression for the surface impedance of the two-layer unbounded waveguide:

$$\xi = -j\frac{Z_0}{ka_1}\frac{1}{F} = \frac{v_1}{\sigma_1}\frac{S + jtg(v_1 d_1)}{tg(v_1 d_1)S - j} \tag{25}$$

For the perfectly conducting external layer $(S = 0)$ and thin internal layer $(tgx \approx x)$ the surface impedance is clearly real and clearly geometrical: $\xi = kZ_0 d_1$ and turns to zero for the perfectly conducting pipe $(d_1 = 0)$.

Returning to (23), we write it for small $S$ $(S << 1)$:

$$\frac{1}{F} = -\frac{ja_1\varepsilon_0 s\, \sec^2(d_1 v_1)}{\varepsilon_1} + \frac{a_1\varepsilon_0 s\, tg(d_1 v_1)}{\varepsilon_1}. \tag{26}$$

Further simplifying the expression (26), we write it for small $d_1$:

$$\frac{1}{F} = \frac{2k^2}{k_d^2} + \frac{1-j}{2} k^{3/2} s_2^{3/2} \tag{27}$$

where $k_d = \omega_d/c$ and $s_2$ defined by (19). The equation for the eigenvalues is now converted to the form:

$$\frac{v_{0,i} a_1 J_0(v_{0,i} a_1)}{J_1(v_{0,i} a_1)} = \frac{2k^2}{k_d^2} + \frac{1-j}{2} k^{3/2} s_2^{3/2} \tag{28}$$

Consider now the longitudinal propagation constant of the mode $TM_{0,i}$:

$$K = \sqrt{k^2 - v_{0,i}^2/a_1^2}. \tag{29}$$

The most obvious possibility of availability of a synchronous component (when the phase velocity of the wave is equal to the velocity of the particle $v$, i.e., $K = k$ at $v = c$) of the mode at a particular frequency is a vanishing of the transverse eigenvalue at a given frequency. As follows from equation (28), in this case, the only way to get the real value of the wavenumber occurs, when $s_2 = 0$, i.e. in the case of perfectly conducting outer layer. Desired wavenumber is $k = k_d$, the geometrical wavenumber, previously determined.

Vanishing of eigenvalue on the resonance frequency determines the uniqueness of solutions of (28) for $s_2 = 0$. It is presented in Figure 7.

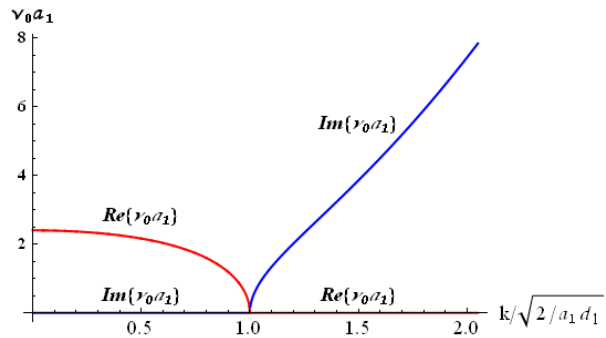

Fig.7. Real (red) and imaginary (blue) parts of transverse eigenvalue for two-layer waveguide with perfectly conducting external layer and thin metallic inner layer for $TM_{01}$ mode.

Eigenvalue, in this case, is either purely real (for $k < k_d$) or purely imaginary (for $k > k_d$).

Another possibility, which is not so obvious, is the equality of the real and imaginary parts of the transverse eigenvalue near the resonant frequency.

In the case of $k \gg |v_{0,i}/a_1|$ the longitudinal propagation constant (29) can be written as:

$$K = k - v_{0,i}^2 / 2ka_1^2. \tag{30}$$

If $v_R = \mathrm{Re}\, v_{0,j}$ and $v_I = \mathrm{Im}\, v_{0,j}$ one obtains for $v_R = v_I = \tilde{v}$:

$$K = k - j \frac{\tilde{v}^2}{ka_1^2}, \tag{31}$$

i.e. we have the synchronous wave with damping rate (decrement) $P = 2v_R v_I / ka_1^2$. As it follows from the equation (28), the module of $v_{0,i}$ should be small for small $s_2$, so let's rewrite (28) for the small $|v_{0,i}|$ values:

$$2 - \frac{a_1^2 v_{0,i}^2}{4} = \frac{2k^2}{k_d^2} + \frac{1-j}{2} k^{3/2} s_2^{3/2} \tag{32}$$

Now we can write the condition of equality of real and imaginary parts of eigenvalue by equating to zero the real terms of equation (32):

$$\frac{2k^2}{k_d^2} + \frac{1}{2} k^{3/2} s_2^{3/2} - 2 = 0 \tag{33}$$

If $s_2 \ll 1$, then $k \approx k_d + dk$ with $dk \ll k$. After some simple calculations and approximations, which we omit, one obtains the expression for the addition to the sinchronized frequency, which coincides with the previous result, obtained for the resonant frequency for $s_2 \ll 1$:

$$dk \approx -\frac{1}{8} k_d^{5/2} s_2^{3/2} \tag{34}$$

Then, equating to zero the imaginary terms of (32) we obtain the expressions for the equal to each other real and imaginary parts of the eigenvalue:

$$a_1^2 v_R^2 = k_d'^{3/2} s_2^{3/2}, \quad v_R = \frac{k_d'^{3/4} s_2^{3/4}}{a_1}, \tag{35}$$

where

$$k_d' = k_d + dk \tag{36}$$

Here $k_d$ is a geometric wavenumber (expressed through the main geometric parameters of the waveguide), and $dk$ is a material amendment.

As shown in Table 2, the small corrections to the basic geometric frequencies are well approximated by the resonance frequencies obtained with the exact solutions. The corresponding distribution, obtained by solving the equation (28) is shown in Figure 8.

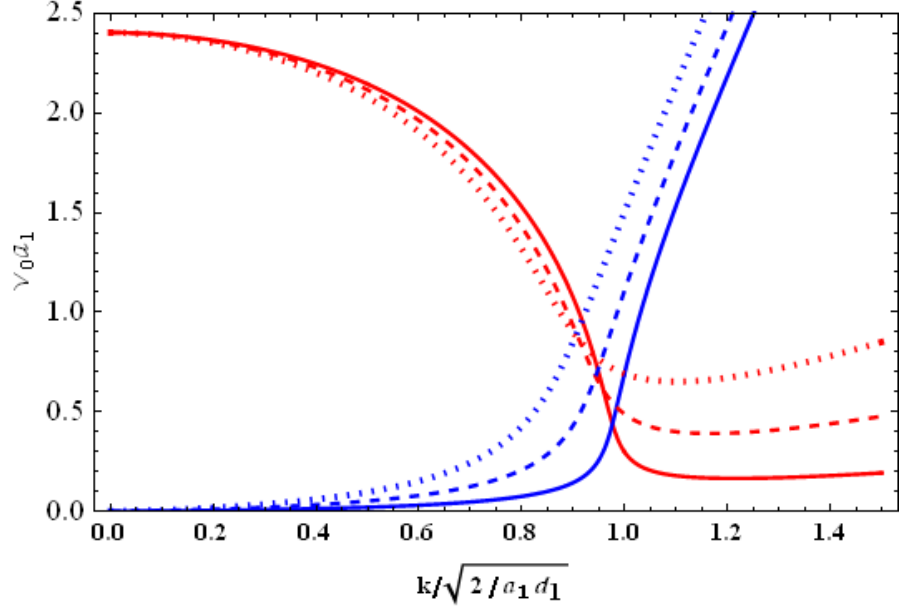

Fig.8. Real (red) and imaginary (blue) parts of transverse eigenvalues of $TM_{01}$ mode of two-layer waveguide with finite conducting external layer; $s_2^{3/2}k_d^{3/2} = 0.2$ (solid), 0.5 (dashed), 0.9 (dotted).

As can be seen, in this case, closer to reality, the eigenvalues are complex values in the entire frequency distribution region. The intersection points of the corresponding red and blue curves representing the equal real and imaginary components of eigenvalues, correspond to the resonance frequencies. In particular, in the case of $a_1 = 1mm$, $\sigma_2 = 58 \cdot 10^6 \, \Omega^{-1} m^{-1}$ and $d_1 = 0.2 \mu m$ (see Figure 2 and Table 1, I, II) the value of $s_2^{3/2}k_d^{3/2}$ is equal to $\approx 0.302$. The numerical solution of (28) for this case is shown in Figure 9. As one can see, the point of intersection of the real and imaginary parts of eigenvalue curves has a small shift to lower frequencies and corresponds to the resonance frequency of 4.6 THz, which coincides with the above-defined frequency with the help of the asymptotic formulae (18), and with the help of numerical calculations as well.

Sequence of point charges or bunches separated by the distance $\Delta z = 2\pi c n / \omega_{rez}$ $(n = 1,2,3...)$, corresponding to the resonance frequency, will strengthen the generation of synchronous mode while suppressing all remaining.

In the limiting case of $d_1 \to 0$, $k_d \to \infty$, the first term on the right-hand side of (28) vanishes and the equation (28) reduces to the case of a single-walled tube. In this case it can be written in a form of dimensionless wavenumber $\kappa = k s_2$;

$$\frac{v_{0,i} a_1 J_0(v_{0,i} a_1)}{J_1(v_{0,i} a_1)} = \frac{1-j}{2} \kappa^{-3/2} \qquad (37)$$

As follows from Figure 10, in this case the intersection of the curves for the real and imaginary parts of the eigenvalue at the maximum of the real part of the impedance $(\kappa = 1.74)$, or in its vicinity, is not happening, contrary to the assertion of [8].

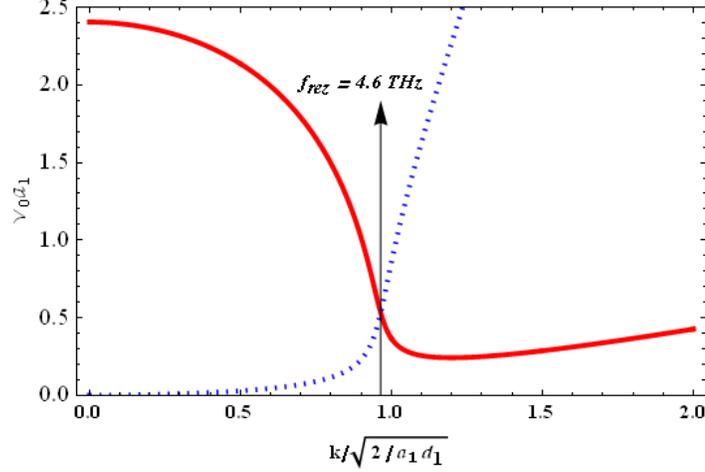

Fig. 9. Real (red) and imaginary (blue) parts of transverse eigenvalues of $TM_{01}$ mode of two-layer waveguide with finite conducting external layer; $s_2^{3/2} k_d^{3/2} = 0.302$.

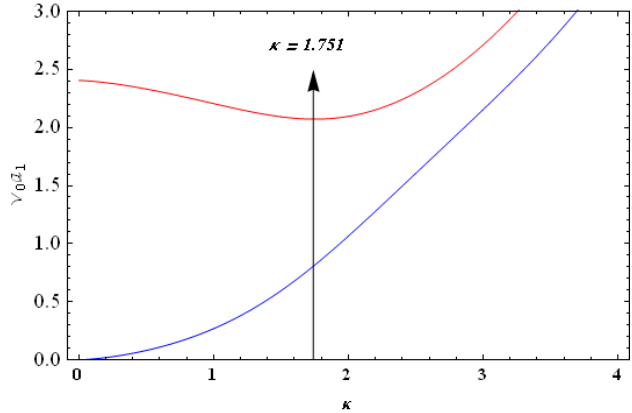

Fig.10. Real (red) and imaginary (blue) parts of transverse eigenvalues of $TM_{01}$ mode of resistive wall single-layer unbounded waveguide.

## 6. Wake functions

Wake function of Gaussian bunch of RMS length $\sigma_s$ can be written as the inverse Fourier transform of impedance with a weight function $\exp(-\omega^2 \sigma_s^2 / 2c^2)$ [9]:

$$W_\parallel(s) = -\frac{1}{2\pi} \int_{-\infty}^{\infty} Z_\parallel(\omega) e^{-j\frac{\omega}{c}s} e^{-\omega^2 \sigma_s^2 / 2c^2} d\omega \qquad (38)$$

In the case of a perfectly conducting outer layer ($S = 0$) integration of (38) in the complex plane is reduced to the calculation of residues in two simple poles:

$$\omega_{1,2} = \left(-j\beta \mp \sqrt{4\omega_d^2 - \beta^2}\right)/2. \qquad (39)$$

Unlike of the case of unbonded single-layer tube [8], the integrand has no cutoff in the complex plane and explicit form of the wake-function can be written as impact of sum of two residues as a resonatory [8] term:

$$W_s^0(s) = -\frac{Z_0 c}{\pi a_1^2} e^{-\frac{\beta}{2c}f(s)-\frac{\sigma_s^2}{2}k_d^2} \cos(f(s)k_\beta) \quad (40)$$

where

$$f(s) = s - \frac{\beta}{2c}\sigma_s^2, \quad k_\beta = \sqrt{k_d^2 - (\beta/2c)^2}, \quad (41)$$

In the case of point-like charge $(\sigma_s = 0)$ expression (41) simplifies to:

$$W_s^0(s) = -\frac{Z_0 c}{\pi a_1^2} e^{-\frac{\beta s}{2c}} \cos(k_\beta s) \quad (42)$$

Note that for the cases of practical interest $k_d \gg \beta/2c$ $(4 - \omega_d^2 \frac{\varepsilon_0^2}{\sigma_1^2} > 0, \; d_1 > \frac{c^2}{2a_1}\frac{\varepsilon_0^2}{\sigma_1^2} = \frac{1}{2a_1\sigma_1^2 Z_0^2})$
and $k_\beta \approx k_d$. In this simplest case, the wake function is given by the periodic curve with decreasing oscillations (Fig.11). The oscillation amplitude decays exponentially and the damping rate of the oscillations is determined by parameter $\beta$ (17).

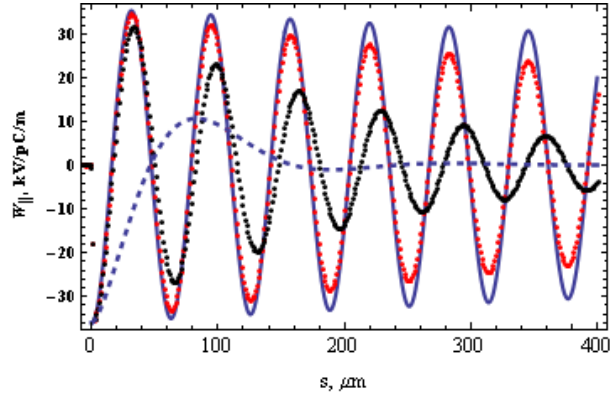

Fig.11. Point-like charge wake function for two-layer waveguide, $a_1 = 1mm$, $d_1 = 0.2\mu m$, $a_3 = 2mm$, $\sigma_1 = 3 \cdot 10^4 \Omega^{-1}m^{-1}$; 1) perfectly conducting external layer (blue, solid), calculated by formula (42); 2) perfectly conducting external layer (red, dotted), calculated by formulae (1),(2); 3) finite conducting external layer (black, dotted), $\sigma_2 = 58 \cdot 10^6 \Omega^{-1}m^{-1}$; 4) single-layer waveguide (blue, dashed), $a_1 = 1mm$, $a_2 = 2mm$, $\sigma_1 = 3 \cdot 10^4 \Omega^{-1}m^{-1}$.

Let's compare the curve for the wake function (Figure 11), constructed from the simplified formula (42) with the curve obtained from numerical calculations by the exact expressions (1), (2).

The simplified formula (42) for the chosen parameters correctly describes the periodicity of the wake function but slows its damping. Attenuation increases with the presence of the finite conductivity of the external layer (black curve). This also increases the oscillation period of the wake function.

Figure 12 shows the dependence of the properties of the wake function from the conductivity of the lower layer. An increase in the conductivity of the bottom layer significantly increases the attenuation of the wake function, but does not acts on its period and the location of the extremes. In all the cases, the value of the wake function on the charge is equal to the same value $W_s^0(s=0) = -\dfrac{Z_0 c}{\pi a_1^2}$ and is independent from the electrodynamic parameters of the layers and from the inner layer thickness as well.

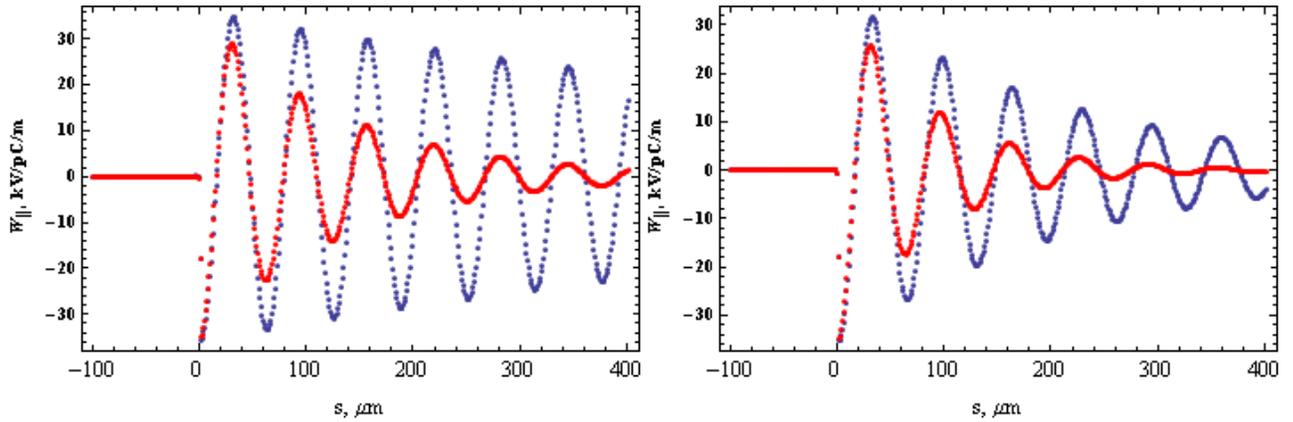

Fig.12. Point-like charge wake function for two-layer waveguide, $a_1 = 1mm$, $d_1 = 0.2\mu m$, $a_3 = 2mm$; $\sigma_1 = 3\cdot 10^4 \Omega^{-1} m^{-1}$ (blue), $\sigma_1 = 3\cdot 10^5 \Omega^{-1} m^{-1}$ (red); perfect conductivity of external layer (left); finite conducting external layer with $\varpi_2 = 58\cdot 10^6 \Omega^{-1} m^{-1}$ (right).

The basic properties of radiation in multi-bunch mode (Fig.13) can be obtained from a superposition of the analytical expressions (42), for point bunches, and (40) for Gaussian bunches.

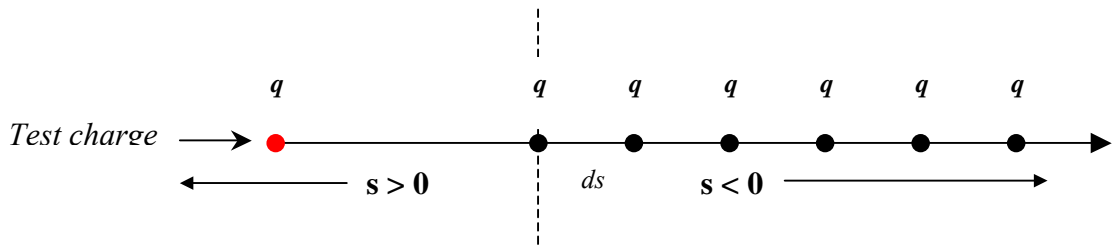

Fig.13. Multi-bunch mode

Sequence of point charges located at the same distance $ds$ from each other (Figure 13) creates, behind him (for $s > 0$, Fig.13), a field that can be expressed as the sum of contributions from each of the bunches.

$$E_z(s) = q\sum_{i=0}^{N} W_s^0(s + i\,ds) = -q\frac{Z_0 c}{\pi a_1^2}\sum_{i=0}^{N} e^{-\frac{\beta(s+ids)}{2c}}\cos(k_\beta s + ik_\beta ds) \qquad (43)$$

Here $q$ is a charge value and $N$ is a number of charges in sequence. If $ds = 2\pi/k_\beta$, then (43) turns to the sum of a geometric progression:

$$E_z(s) = -q\frac{Z_0 c}{\pi a_1^2}\mathrm{H}e^{-\beta s/2c}\cos(k_\beta s) \qquad (44)$$

where

$$\mathrm{H} = \frac{1 - \exp(-N\beta\,ds/2c)}{1 - \exp(-\beta\,ds/2c)} \qquad (45)$$

Thus, the structure of the field behind the synchronized $(ds = 2\pi/k_\beta)$ sequence of bunches is identical to the wake field of a single bunch and H - form-factor characterizes the growth of coherent radiation as a result of interaction of the bunches. For $N \gg 1$ H factor can be expressed as the sum of an infinite geometric series: $\mathrm{H} = \{1 - \exp(-\beta\,ds/2c)\}^{-1}$. For $a_1 = 1mm$, $d_1 = 0.2\mu m$, $\sigma_1 = 3\cdot 10^4 \Omega^{-1}m^{-1}$ and $N \to \infty$ $\mathrm{H} = 36.5$. Field (longitudinal electric component) in the cross-section plane of the last bunch ($s \to 0$) is represented as

$$E_z(s \to 0^+) = -q\frac{Z_0 c}{\pi a_1^2}\mathrm{H} \qquad (46)$$

For the parameters mentioned above, this value is $1.3\cdot 10^6 V/pC/m$.

## 7. Radiation from the open end of waveguide.

Suppose that the waveguide is semi-infinite and cross-sectional area of the aperture is located at $z = 0$. Not including edge effects one can assume the field incident on the open end of the waveguide undisturbed, ie the same as in the infinite waveguide [10]. Since the wakefield, incident on the aperture of the waveguide is not entirely monochromatic, one should first calculates the emission of its frequency component from the aperture [11]. Distribution of the single transverse (radial) electrical component of wakefield in the frequency domain, running on the open end of the waveguide $(z = 0)$ after the release of the last bunch is equal to

$$E_r(r,\omega) = -\mathrm{H}\frac{kr}{2}\frac{Z_0 c}{\pi a_1^2}(X + jY)^{-1}e^{-j\omega t}, \quad 0 \le r \le a \qquad (47)$$

with $S = 0$. The far-field radiation is reconstructed with the help of the distribution (47) in the aperture according to work [12]:

$$\vec{E}(\vec{R}) = k\cos\vartheta \vec{A}(k_x, k_y)\frac{e^{ikR}}{R} \qquad (48)$$

where

$$A_{x,y}(k_x,k_y) = \int_0^{2\pi}\int_0^a E_{x,y}(x,y,0)e^{-ik\sin\theta\cos(\varphi-\alpha)}rdrd\alpha, \quad A_z(k_x,k_y) = -\frac{1}{k_z}\{k_x A_x(k_x,k_y) + k_y A_y(k_x,k_y)\} \quad (49)$$

and

$$k_x = k\sin\theta\cos\phi, \ k_y = k\sin\vartheta\sin\phi, \ k_z = \sqrt{k^2 - k_x^2 - k_y^2}$$
$$E_x = E_r(r,\omega)\cos\alpha, \ E_y = E_r(r,\omega)\sin\alpha \quad (50)$$

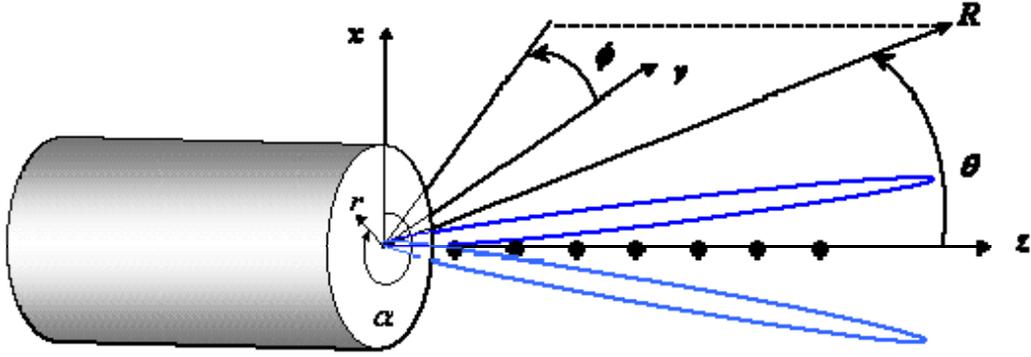

Fig.14. Radiation from the open end of round waveguide.

In (48) - (50) $r,\alpha$ are the polar coordinates associated with the center of the waveguide opening, $R,\theta,\varphi$ are the spherical coordinates of the observation point, located in the far zone (Fig. 14). After performing the appropriate integrations and substitutions, we obtain for the spectral density of far-field:

$$E_\varphi(\vec{R},\omega)d\omega = -jq\mathrm{H}kZ_0 c(X+jY)^{-1}\frac{J_2(ka_1\sin\theta)}{\sin\theta}\cos\theta\sin\varphi e^{-j\omega t}\frac{e^{jkR}}{R}d\omega$$
$$E_\theta(\vec{R},\omega)d\omega = jq\mathrm{H}kZ_0 c(X+jY)^{-1}\frac{J_2(ka_1\sin\theta)}{\sin\theta}\cos\varphi e^{-j\omega t}\frac{e^{jkR}}{R}d\omega, \quad E_R = 0. \quad (51)$$

The space-time representation of the field is obtained by integrating the spectral distribution (51) on frequency:

$$\vec{E}(\vec{R}) = \int_{-\infty}^{\infty}\vec{E}(\vec{R},\omega)d\omega \quad (52)$$

Applying the technique of integration on the complex plane, we calculate the contributions of residues at the poles (39). The result is:

$$\begin{Bmatrix} E_\varphi(\vec{R}) \\ E_\theta(\vec{R}) \end{Bmatrix} = -jq\mathrm{H}Z_0 \sum_{1,2} \omega_{1,2} \frac{J_2\left(\frac{\omega_{1,2}}{c} a_1 \sin\theta\right)}{\sin\theta} \begin{Bmatrix} \cos\theta\sin\varphi \\ \cos\varphi \end{Bmatrix} e^{-j\omega_{1,2}t} \frac{e^{j\frac{\omega_{1,2}}{c}R}}{R} \qquad (53)$$

For small $\beta$ (compared to $\omega_d$) can be approximately written (for $t \geq R/c$):

$$\begin{Bmatrix} E_\varphi(\vec{R}) \\ E_\theta(\vec{R}) \end{Bmatrix} = 2q\mathrm{H}Z_0 \frac{c}{d_1} F(\theta) e^{-\frac{\beta}{2}(t-R/c)} \frac{e^{j\omega_\beta(t-R/c)}}{R} \begin{Bmatrix} \cos\theta\sin\varphi \\ \cos\varphi \end{Bmatrix} \qquad (54)$$

with

$$F(\theta) = \frac{J_2\left(\sqrt{2a_1/d_1}\sin\theta\right)}{\sqrt{2a_1/d_1}\sin\theta} \qquad (55)$$

The maximum power radiated in the front half-space for $a_1 \gg d_1$ is defined by the expression

$$P \approx \frac{\pi q^2 Z_0 \omega_d^2}{2}\left(1 - \frac{d_1}{2a_1}\right)\mathrm{H}^2 e^{-\beta(t-R/c)} \qquad (56)$$

For $a_1 = 1mm$, $d_1 = 0.2\mu m$, $\sigma_1 = 3\cdot10^4 \Omega^{-1}m^{-1}$ for the single charge $(\mathrm{H}=1)$ with $q = 1pC$ the maximum radiated power $(t = R/c)$ is equal to $0.5MW$. A tenfold increase in the radius of the waveguide $(a_1 = 1cm)$ leads to a tenfold reduction in radiated power $(\sim 50kW)$.

The radiation pattern is zero at the principal direction $\theta = 0$ and reaches its maximum value at $\theta = \arcsin\left(2.3\sqrt{d_1/2a_1}\right)$. Meanwhile, at $d_1/2a_1 \ll 1$ it is close to the principal direction. The main output power (about 75%) is contained in a cone with a resolution $\Delta\theta = 2\arcsin\left(\chi\sqrt{d_1/2a_1}\right)$, where $\chi = 5.13562$ the first root of the Bessel function $J_2(x)$. The increase of the inner radius of waveguide can narrow the directional lobes (Figure 15) and keeping the percentage of radiated power in the main lobe.

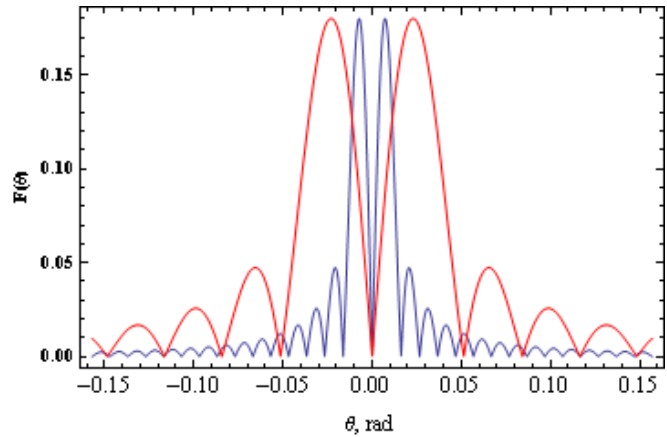

Fig.15. Radiation pattern, $d_1 = 0.2\mu m$; $a_1 = 1mm$, $f_d = 4.775THz$ (red); $a_1 = 1cm$ (blue) $f_d = 1.510THz$.

The angular distribution of the normalized pattern is uniquely dependent on the ratio $a_1/d_1$. Increase in this ratio leads to a narrowing of the radiation pattern, at the same time reducing the operating frequency, which also depends mainly on the geometrical parameters of the waveguide $\omega_d = c\sqrt{2/a_1 d_1}$.

On the film, located at some distance from the radiating open end of the waveguide parallel to the plane of its aperture, will engrave image of the ring. Maximum blackening on a screen located at a distance of 1m, for $a_1 = 1mm$ and $d_1 = 0.2\mu m$ should be ring-shaped radius $23mm$ and for $a_1 = 1cm$ and $d_1 = 0.2\mu m$ the ring radius should be about to $7.3mm$.

## 8. Two-beam acceleration

We place the test point charge $q_t$ at a distance $s_t = \pi/k_d$ from the last point charge of the same sequence of bunches, in the point of maximum accelerating wakefield. The field accelerating of the test charge, in the mentioned point is equal to

$$E_z(s_t) = q\frac{Z_0 c}{\pi a_1^2} \mathrm{H} e^{-\beta s_t/2c}, \tag{57}$$

whereas the maximum of decelerating field, acting on driving sequence of charges (at $s = 0$), is equal to

$$E_z(s=0) = -q\frac{Z_0 c}{2\pi a_1^2} \mathrm{H} \tag{58}$$

Thus, the transformation ratio is equal to $R = |E_z(s_t)/E_z(s=0)| = 2e^{-\beta s_t/2c} < 2$, which is consistent with the theorem on collinear two-beam acceleration (wakefield theorem) [13,14]. Increasing of transformation ratio is achieved in various ways, mainly connected with the introduction of inhomogeneities in the charge distribution in the bunches or a variation of the charge in the sequences of bunches in the case of collinear driving and witness bunches sequences [15-18], or transmission the energy, generated by the leading bunch to the parallel structure, designed for the witness bunch for non-collinear cases [19, 20]. However, as will be shown below, even using the described simple scheme of two-beam collinear acceleration, can be obtained a significant rate of acceleration.

The energy gain of accelerated beam (per meter) can be defined by the formula

$$\frac{d\Sigma}{cdt} = q_t E_z(s_t) = q_t q \frac{Z_0 c}{\pi a_1^2} \mathrm{H} e^{-\beta s_t/2c} \tag{59}$$

In particular, the driving train of 20 particles with a charge of 1 nC in the waveguide of radius $a_1 = 1mm$ ($d_1 = 0.2\mu m$, $\sigma_1 = 3\cdot 10^4 \Omega^{-1} m^{-1}$) provides the acceleration rate of the order of $0.55 GeV/m$. The distance between the particles (bunches) in the driving train is equal to $62.8\mu m$, and the train of 20 particles has a length about $1.25mm$. Waveguide length of 2.5 m provides acceleration about of $1 GeV$. The smallness of the transformation ratio ($< 2$) is compensated by a large absolute value of the energy, released as a result of deceleration of the leading bunches. Coherence of wakefield radiation provides its effective interaction with the accelerated bunch.

Experimental verification of the presence of accelerated electrons can be produced by a magnetic spectrometer (the main part of which is the dipole magnet), deflecting the electrons at different angles depending on their energy. The same principle can be used to isolate the accelerated bunches from the total flow, which consists from the driving and accelerated bunches.

## 9. Conclusion

Our studies suggest the possibility of obtaining a narrow-band radiation in the terahertz frequency range, generated by a bunch or a sequence of bunches of charged particles during of passage along of the axis of the two-layered metallic (resistive) waveguide. At sufficiently high conductivity of the metal filling the outer layer, the resonance frequency region depends mainly on the inverse square of inner radius of the waveguide and the thickness of inner coating. By varying the inner radius of the waveguide from 1 cm to 1 mm the resonant radiation at frequencies from 0.5 to 6.5 THz can be obtained. Obviously, the smaller the radius of the waveguide, the higher is the resonant frequency (for the same thickness of the inner layer). For fixed geometrical dimensions of the waveguide the upper frequency limit for the resonance frequency is the geometrical resonant frequency $\omega_d = c\sqrt{2/a_1 d_1}$, valid for a perfectly conducting outer layer. Finite conductivity of the material of the external layer may to slightly lower (no more than 10%) the value of resonant frequency. According to the formula for $\omega_d$, increase the resonance frequency can be achieved by thinning of the lower layer. However, as shown above (see Figs 2 and 3), too thin inner layer is no longer a source of resonance. For a given combination of the conductivities of the external and internal layers there is a thickness of inner layer, at which the resonance phenomenon is most pronounced in meaning of the level of resonance and the width of the resonance curve. Thus, when the inner radius of the waveguide is 1mm, for $\sigma_2 = 58 \cdot 10^6 \Omega^{-1} m^{-1}$ (copper) and $\sigma_1 = 3 \cdot 10^4 \Omega^{-1} m^{-1}$ (NEG) most characteristic resonance occurs at 2.97 THz, with thickness of the inner layer of 0.5 μm (see Fig. 2, left and Tab. 1, I) and when the inner radius of the waveguide is 1cm, for $\sigma_2 = 58 \cdot 10^6 \Omega^{-1} m^{-1}$ (copper) and $\sigma_1 = 3 \cdot 10^5 \Omega^{-1} m^{-1}$ (NEG) most characteristic resonance occurs at 0.97 THz, with thickness of the inner layer of 0.5 mm (see Fig. 2, right and Tab. 1, IV).

Analytical and numerical calculations performed for the source of terahertz radiation (Chapter 7) and for the two-beam acceleration (Chapter 8) are satisfied for the point charge in the case of perfectly conducting upper layer of the waveguide. Finite conductivity of the upper layer and the finite length of the bunch will lead to some erosion of the annular image of far-field radiation and in some reduction in the rate of acceleration, but the basic pattern remains the same.

## 10. References.

[1] http://xfel.desy.de/technical_information/tdr/tdr/
[2] http://www.psi.ch/swissfel/
[3] J. B. Rosenzweig, G. Andonian, P. Muggli, P. Niknejadi, G. Travish et al. AIP Conf. Proc. 1299, 364 (2010)
[4] K.L.F. Bane and G. Stupakov, SLAC-PUB-14839, December 2011
[5] M.Ivanyan, A.Tsakaniam, Proc. of IPAC2011, San Sebastián, Spain, pp. 703-705, 2011
[6] M. Ivanyan, V. Tsakanov, Phys. Rev. ST-AB, 7, 114402 (2004).
[7] M.Ivanyan, E.Laziev, A.Tsakanian, V.Tsakanov, A.Vardanyan, S.Heifets, Phys. Rev. ST-AB, 11, 084001 (2008).
[8] K. L. F. Bane and M. Sands, Report No. SLAC-PUB-95-7074, 1995

[9] B.W. Zotter and S.A. Kheifetz, Impedances and Wakes in High-Energy Particle Accelerators, World Scientific, Singapore, 1997.
[10] L.A. Vainshtein, The theory of diffraction and the factorization method, Sovetskoe Radio, Moscow, 1966 (in Russian).
[11] M. Born and E. Wolf, Principles of Optics, 6th ed., Pergamon, Oxford, 1980.
[12] Joy E.B., Paris D.T., IEEE Trans. AP-20, 3, pp. 253-261, 1972.
[13] P.B. Wilson, in Proc. of the 13th SLAC Summer Institute On Particle Physics (SLAC Report No. 296), pp. 273–295.
[14] J.G. Power, W. Gai, and P. Schoessow, Phys. Rev. E 60, 6061 (1999).
[15] K. L. Bane, P. Chen, and P.B. Wilson, IEEE Trans. Nucl.Sci. 32, 3524 (1985).
[16] P. Schutt, T. Weiland, and V.M. Tsakanov, in Proc. of the Nor Amberd Conference, Armenia, Vol. 7, p. 12 (1989)
[17] A.G. Ruggiero, P. Schoessow, and J. Simpson, in Proc. of the Advanced Accelerator Concepts: 3$^{rd}$ Workshop, edited by F. E. Mills (American Institute of Physics, Madison, WI, 1986), p. 247.
[18] C. Jing, A. Kanareykin, J.G. Power, M. Conde, Z. Yusof, P. Schoessow, and W. Gai, Phys. Rev. Lett. 98, 144801 (2007).
[19] Wanming Liu and Wei Gai, Phys. Rev. ST-AB, 12, 051301 (2009).
[20] Wei Gai, Manoel Conde, John Gorham Power, Proc. of IPAC10, Kyoto, Japan, pp. 3428-3420 (2010).